\documentclass[epsfig,amsfonts]{mn}

\usepackage{graphicx}
\usepackage{amsfonts}
\usepackage{amssymb}

\title{3D Weak Gravitational Lensing of the CMB and Galaxies}
\author[Kitching, Heavens, Das]{T. D. Kitching$^1$\thanks{t.kitching@ucl.ac.uk}, A. F. Heavens$^2$, S. Das$^3$\\ 
$^1$Mullard Space Science Laboratory, University College London, Holmbury St Mary, Dorking, Surrey RH5 6NT, UK\\
$^2$Imperial Centre for Inference and Cosmology, Imperial College London, Prince Consort Road, London SW7 2AZ, UK\\
$^3$Berkeley Center for Cosmological Physics, Dept. of Physics, University of California, Berkeley, CA, USA 94720}

\newcommand{\be}{\begin{equation}}  \newcommand{\ee}{\end{equation}}
  \newcommand{\ba}{\begin{eqnarray}}
\newcommand{\ea}{\end{eqnarray}}  
\newcommand{\nn}{\nonumber\\}

\newcommand{\bD}{{\bf D}}

  \newcommand{\br}{{\bf r}}

\def\gs{\mathrel{\raise1.16pt\hbox{$>$}\kern-7.0pt %
\lower3.06pt\hbox{{$\scriptstyle \sim$}}}}         %
\def\ls{\mathrel{\raise1.16pt\hbox{$<$}\kern-7.0pt %
\lower3.06pt\hbox{{$\scriptstyle \sim$}}}}         %

\newcommand{\apj}{Astrophysical Journal }
\newcommand{\apjl}{Astrophysical Journal Letters}
\newcommand{\apjs}{Astrophysical Journal Supplements}

\newcommand{\mnras}{Mon.Not.Roy.Astron.Soc. }

\newcommand{\prd}{Physical Review D }

\newcommand{\physrep}{Physics Reviews}
\newcommand{\procspie}{Proceedings of SPIE}

\begin{document}

\voffset=-0.25in 

\maketitle

\begin{abstract}
In this paper we present a power spectrum formalism that combines the full 
three-dimensional information from the galaxy ellipticity field, 
with information from the cosmic microwave background (CMB). 
We include in this approach galaxy cosmic shear and galaxy intrinsic alignments, 
CMB deflection, CMB temperature and CMB polarisation data; including the inter-datum power spectra between all quantities. 
We apply this to forecasting cosmological parameter errors for CMB and imaging surveys 
for Euclid-like, Planck, ACTPoL, and CoRE-like experiments. 
We show that 
the additional covariance between the CMB and ellipticity measurements can improve 
dark energy equation of state measurements by 15\%, 
and the combination of cosmic shear and the CMB, from Euclid-like and CoRE-like experiments, 
could in principle measure the sum of neutrino masses with an error of $0.003$ eV.
\end{abstract}


\section{Introduction}
Observations of the cosmic microwave background (CMB) have been used to infer cosmological parameter values with unprecedented 
accuracy and precision, most recently with the \emph{Planck} (Planck Collaboration, 2013a) results. These 
measurements have helped to establish the currently favoured cosmological paradigm: a universe with a flat geometry, 
dominated by dark matter and dark energy. However the CMB is limited in its ability to measure low-redshift phenomena, 
for example the transition from a dark matter dominated epoch to a dark energy dominated epoch, because it provides only a single 
source redshift of photons that probe the physics at the surface of last scattering and the integrated effect of the expansion 
history and growth along the line of sight. 

To probe the low-redshift physics, that governs the transition from dark matter to dark energy domination, requires cosmological 
observations that provide many data points that sample this era. One such probe is cosmic shear, that uses the weak lensing information 
imprinted on galaxy images. Galaxy weak lensing is the effect where gravitational lensing, caused by matter perturbations along the line of sight,  
causes a change in the observed third flattening (or third eccentricity) or observed size of 
galaxy images. The change in third flattening is colloquially referred to as `ellipticity', and the 
additional ellipticity caused by lensing known as `shear'; in this paper we only look at changes in the ellipticity 
not the size of galaxies, we refer the reader to Heavens, Alsing, Jaffe (2013) for a discussion of size changes caused by weak lensing.
By measuring the ellipticity of many galaxies, and calculating the variance of the ellipticity as 
a function of scale, cosmological information can be extracted through the dependency of this statistic on the 
power spectrum of matter perturbations, the expansion history of the Universe, and the growth of structure. 
Because each galaxy is observed at a particular angular coordinate on the sky, and with a particular redshift estimate, 
the shear information from a population of galaxies is naturally described using 3D analyses (Heavens, 2003). The analysis of shear in 3D for the purposes of 
cosmology is known as 3D cosmic shear. This has been shown to be a particularly good method for determining dark energy (Kitching, 2007), modified gravity 
(Heavens, Kitching, Verde, 2007) and neutrino mass parameters (Kitching et al., 2008; Jimenez et al., 2010). 

In this paper we show how CMB and 3D cosmic shear information can be combined in a single formalism that uses a spherical-Bessel harmonic transform. In doing this we 
account for the weak lensing of the CMB, which causes a non-zero covariance between 3D cosmic shear and the CMB -- both being lensed by the same large-scale structure. 
Lensing of the CMB has been detected in a series of experiments (for example Planck Collaboration, 2013b; Das et al., 2014; van Engelen et al., 2014) and a 
cross-correlation between galaxy lensing and 
CMB lensing has also been detected at $\approx 3$-sigma (Hand et al., 2013). 

We also generalise the 3D cosmic shear formalism to include the possible correlations between the 3D shear field and the unlensed `intrinsic' ellipticity 
field of galaxies. The `intrinsic alignment' of galaxies (see Troxel \& Ishak, 2014a for a recent review) is a potential systematic effect for 3D cosmic shear because 
the intrinsic alignment power spectrum can mimic the cosmological signal (e.g. Heymans et al., 2013). The investigation of calibrating intrinsic alignments by including
CMB lensing has also been studied by Hall \& Taylor (2014) and Troxel \& Ishak (2014b) who both looked at the impact of intrinsic alignments on 
coarsely binned 2D cosmic shear power spectra; here we present a fully 3D analysis and also propagate the investigation through to predicted cosmological parameter errors.  
 
Such combinations of data are commonly referred to as `cross-correlations', a term that refers to the act of finding inter-datum combinations that may 
contain additional information beyond a simple combination of the parameters' final probabilities from the individual data sets. We avoid such 
terminology here, and refer to inter-datum combinations and intra-datum combinations to avoid reference to a data analysis that would involve the computation of 
any correlation function. In the approach we present, the data vector from an individual experiment 
would be supplemented with the data vector from another, and 
the theoretical covariance of this combined data vector - which contains the cosmological information in this case - now needs to include the extra inter-datum covariance between the data vectors as well as the intra-datum covariance of the original data vectors. 

We present the formalism in Section \ref{Methodology}. In Section \ref{Results} we show predictions for cosmological parameter 
constraints. We discuss conclusions in Section \ref{Conclusion}. 

\section{Methodology}
\label{Methodology}
In this Section we present the general formalism for combining 3D cosmic shear power spectra and CMB lensing power spectra, 
importantly we derive the cross-power term. We refer the reader to Kitching et al. (2014) for a detailed discussion of the 3D cosmic shear formalism, and an 
application to data, and to Lewis \& Challinor (2006) for a detailed discussion of CMB weak lensing.

The data vector with which we are concerned is the combination of the observed galaxy ellipticities, 
and the CMB temperature and polarisation measurements. We can write this as 
\be 
\bD=\{e, T, p\}
\ee
where $e$ is the measured galaxy ellipticity of an object, a spin-2 quantity $e=e_1+{\rm i}e_2$, the CMB polarisation $p=|p|{\rm exp}(2{\rm i}\theta_p)$ is also a spin-2 quantity typically 
assigned an amplitude $|p|$ and an angle of polarisation $\theta_p$, 
and the CMB temperature $T$ is a scalar field. The ellipticity is the measured galaxy ellipticity, which is a combination of the galaxies 
unlensed `intrinsic' ellipticity $e^I$ and the additional shear $\gamma$. For small shear 
\be
e\simeq e^I+\gamma, 
\ee
where all the above quantities are spin-2. 

The data vector $\bD$ is observed at galaxy positions $e(r[z],\btheta)$ with 3D coordinate $(r[z],\btheta)$, and at all points 
of the sky for which polarisation and/or temperature data from the CMB are observed. However by taking the spherical harmonic and spherical-Bessel transforms of this data vector, for 
the CMB and 3D cosmic shear parts respectively, we can define a data vector that is continuous in wavenumber and consists of the transform coefficients. Furthermore the 
CMB polarisation data can be use to construct two scalar $E$ and $B$ measurements (see Lewis \& Challinor, 2006; Zaldarriaga \& Seljak, 1997), 
and these in combination with the temperature field can be used to infer a CMB weak lensing deflection 
field $d$ (a spin-1 quantity); $d$ can be derived using a quadratic estimator, for example Hu \& Okamoto (2002). 
The galaxy ellipticity can also be transformed into an $E$ and $B$ mode (see Kitching et al., 2014 Appendix A), 
such that we can write the data vector of the transform coefficients as 
\be
\bD_{\ell,m}(k)=\{e^E_{\ell,m}(k), e^B_{\ell,m}(k), d^E_{\ell,m}, d^B_{\ell,m}, a_{\ell,m},  p^E_{\ell, m}, p^B_{\ell, m}\}
\ee
where $(\ell, m)$ are angular wavenumbers and $k$ is a radial wavenumber. The covariance of these transform coefficients define the power spectrum for each one. 
For example $\langle a_{\ell,m}a^*_{\ell',m'}\rangle=C^{TT}_{\ell}\delta_{mm'}\delta_{\ell\ell'}$, 
$\langle p^E_{\ell,m}p^{E,*}_{\ell',m'}\rangle=C^{EE}_{\ell}\delta_{mm'}\delta_{\ell\ell'}$, 
and $\langle e_{\ell,m}(k)e^*_{\ell',m'}(k')\rangle=C^{ee}_{\ell}(k,k')\delta_{mm'}\delta_{\ell\ell'}$ etc. 
We will label power spectra $C^{XY}_{\ell}$ where $X$ and $Y$ are 
the parts of the data vector between which the covariance is computed. 

In this paper we will simplify the analysis by assuming a flat-sky approximation, and that the galaxy ellipticity and deflection B-modes are 
consistent with noise, such that the data vector that contains cosmological information is 
\be
\bD_{\ell}(k)=\{e^E_{\ell}(k), d^E_{\ell}, a_{\ell},  p^E_{\ell}, p^B_{\ell}\}. 
\ee
Note that this is a complex data vector where the ellipticity and deflection field coefficients are complex quantities. 
We use the symbol $\ell$ to refer 
to the wavevector on the sky, and also the amplitude of the vector in the power spectra. 
For the 3D cosmic shear and deflection field the power spectra used are all E-mode, related to the underlying potentials through 
their respective complex derivatives. The theoretical 
covariances are therefore all E-mode by definition (a B-mode would arise if an imaginary field were also included). 
For data analysis (that contains E and B-mode contributions due to noise) a separation must be made, as described 
in Kitching et al. (2014) appendix A for 3D cosmic shear. We refer to Kitching et al. (2014) for 
an in-depth discussion of this point. For a discussion of the dependency of the modes of the harmonic expansion for 3D cosmic 
shear we refer the reader to Castro, Heavens, Kitching (2005).

A Gaussian likelihood for the full data vector can be written as 
\be
L = \prod_{\ell}\frac{1}{\pi^2|A_{\ell}|^{1/2}}{\rm exp}\left[-\frac{1}{2}\sum_{k_1 k_2}Z_{\ell}(k_1)A^{-1}_{\ell}(k_1,k_2)Z_{\ell}(k_2)\right]
\ee
where the vector $Z_{\ell}(k)=(\bD_{\ell}(k), \bD^*_{\ell}(k))^T$ is a combination of the complex and conjugate parts of the data vector and the covariance matrix $A$ 
accounts for the correlation between the real and imaginary parts of the data vector 
\be
\label{affix}
A_{\ell}(k_1,k_2)=\left( \begin{array}{cc}
\Gamma & R \\
R^T & \Gamma^* \\
 \end{array} \right)
\ee
where the covariance matrix $\Gamma$ and the relation matrix $R$ are related to the covariance matrix of the individual elements of the data vectors 
$C_{\ell}(k_1,k_2)$ by $\Gamma_{\ell}(k_1,k_2)={\mathcal R}[C_{\ell}(k_1,k_2)]+{\mathcal I}[C_{\ell}(k_1,k_2)]$ and 
$R_{\ell}(k_1,k_2)={\mathcal R}[C_{\ell}(k_1,k_2)]-{\mathcal I}[C_{\ell}(k_1,k_2)]$, where we have labelled the parts of the data vector covariance associated 
with the real and imaginary parts with ${\mathcal R}$ and ${\mathcal I}$ respectively. This matrix also includes the respective noise terms for 
each quantity. This is the affix-covariance defined in Kitching et al., (2014), but 
generalised for the extended data vector considered here. 

The covariance matrix of this data vector consists of the inter-datum and intra-datum covariances: 
\be 
\label{pict}
\left( \begin{array}{c|c|c|c|c}
ee & de & Te & Ee & Be \\
\hline
e d & dd & Td & Ed & Bd \\
\hline
eT & dT & TT & ET & BT \\
\hline
eE & dE & TE & EE & BE \\
\hline
eB & dB & TB & EB & BB \\
 \end{array} \right).
\ee
This matrix shows the dependencies in a pictographic manner. We assume in this study that there are no parity-violating modes such that 
the sub-matrices $BE$ and $EB$ are zero. 
On the scales of interest in this paper ($\ell \gg 100$) we ignore the correlation between $T$ and lensing (and $E$ and lensing) 
due to the correlation between the ISW effect and lensing. This means that $Te$, $eT$, $Ee$,
$eE$, $Be$ and $Be$ are zero i.e. that all the lensing information in the CMB is captured in a single inferred deflection field $d$. 
In the absence of these correlations, the lensed CMB fields are uncorrelated with the lensing fields.

This results in the covariance $C_{\ell}(k_1,k_2)$
\ba
\label{signal}
C_{\ell}(k_1,k_2)=
\left( \begin{array}{c|c|c|c|c}
C^{ee}_{\ell}(k_1,k_2) & C^{de}_{\ell}(k_1) & 0 & 0 & 0 \\
\hline
C^{ed}_{\ell}(k_2) & C^{dd}_{\ell} & C^{Td}_{\ell} &  C^{Ed}_{\ell} &  C^{Bd}_{\ell} \\
\hline
0 &  C^{dT}_{\ell} &  C^{TT}_{\ell} &  C^{ET}_{\ell} &  C^{BT}_{\ell} \\
\hline
0 & C^{dE}_{\ell} & C^{TE}_{\ell} & C^{EE}_{\ell} & 0 \\
\hline
0 & C^{dB}_{\ell} & C^{TB}_{\ell} & 0 & C^{BB}_{\ell} \\
 \end{array} \right).\nonumber
\ea
\be
\ee
The sub-matrices that depend on $d$, $T$, $E$ and $B$ depend on angle only (or spherical harmonic transform variable $\ell$). 
The quantities that depend on ellipticity $e$ introduce radial dependence such that for any 
given $\ell$-mode the power spectrum is a $(N_k+4)\times (N_k+4)$ matrix in the $k_1$ and $k_2$ directions, where $N_k$ is 
the number of $k$-modes used.

\subsection{The shear, intrinsic and deflection 3D power spectra}
We can now write down expressions for each of these power spectrum by appealing to the formalism of Heavens (2003) and Kitching,
Heavens, Miller (2011). For the ellipticity-ellipticity power spectra we will decompose this into 
the shear-shear and intrinsic-intrinsic parts; the observed ellipticity being the sum of the two.

\subsubsection{Shear}
For the shear-shear term the theoretical shear transform coefficients are related to the matter over density by (Heavens et al., 2006) 
\ba
\label{shear}
\gamma_{\ell}(k)&=&-D_{\gamma}\frac{3\Omega_{\rm M}H_0^2}{2\pi c^2}\int {\rm d}z_p{\rm d}z' j_{\ell}(kr[z_p])n(z_p)p(z'|z_p)\nn
          &&\int_0^{r[z']}d{\rm r'}\frac{F_K(r,r')}{a(r')}\int {\rm d}k' j_{\ell}(k' r')\frac{\delta_{\ell}(k')}{k'}
\ea
where $F_K=S_K(r-r')/S_K(r)/S_K(r')$ is the lensing kernel where $S_K(r)={\rm sinh}(r),r,\sin(r)$ for cosmologies with
spatial curvature $K=-1,0,1$, $a(r)$ is the dimensionless scale factor at the cosmic time related to the look-back time at 
comoving distance $r$, 
$n(z_p){\rm d}z_p$ is the number of galaxies in a spherical shell of radius $z_p$ and thickness ${\rm d}z_p$,
$p(z'|z_p)$ is the probability of a galaxy at redshift $z'$ to have a photometric redshift $z_p$, $j_{\ell}(kr)$ are spherical Bessel functions, 
$\Omega_{\rm M}$ is the ratio of the total matter density to the critical density at redshift z=0, and $H_0$ is the current value of the 
Hubble parameter.
$\delta_{\ell}(k)$ is the spherical-Bessel transform of the matter over-density field. 
One can also use a facultative factor of $k$ in the transform (as used in Castro, Heavens, Kitching, 2005)                         
but results are unchanged.
The factor $D_{\gamma}=D_{\gamma,1}+{\rm i}D_{\gamma,2}=\frac{1}{2}(\ell_x^2-\ell_y^2)+{\rm i}\ell_x\ell_y$
relates to real and imaginary parts of
the derivative of ${\rm e}^{{\rm i}\ell.\theta}$ with respect to $\theta$ that we show explicitly here; the shear field
being related to the derivative of the lensing potential $\phi$ by
\be
\gamma(\br)=\frac{1}{2}\eth\eth\phi(\br)
\ee
where $\eth=\partial_x+{\rm i}\partial_y$.

The covariance of this expression gives the 3D cosmic shear power spectrum 
\be 
\label{shearpower}
C^{\gamma\gamma}_{\ell}(k_1,k_2)=[D_{\gamma}D^*_{\gamma}]{\mathcal A}^2\int\frac{{\rm d}k'}{k'^2} G^{\gamma}_{\ell}(k_1,k')G^{\gamma}_{\ell}(k_2,k')
\ee
where ${\mathcal A}=3\Omega_{\rm M}H_0^2/\pi c^2$. 
The matrix $G$ is defined as  
\be 
G^{\gamma}_{\ell}(k_1,k')=\int {\rm d}z_p{\rm d}z' j_{\ell}(k_1r[z_p])n(z_p)p(z'|z_p)U_{\ell}(r[z'],k')
\ee
in the continuous limit (i.e. not summing over individual galaxies; see Kitching, Heavens, Miller, 2011). 
The matrix $U$ is an integral over the matter power spectrum and angular diameter distances
\be
U_{\ell}(r[z],k)=\int_0^{r[z]}d{\rm r'}\frac{F_K(r,r')}{a(r')}j_{\ell}(k r')P^{1/2}(k; r'), 
\ee
where $P(k; r)$ is the 
matter power spectrum at comoving distance $r$ at radial wavenumber $k$; we refer the reader to Castro, Heavens, Kitching (2005) for a discussion 
of the approximation involved in using the square-root of the power spectrum here. The lensing potential 
is related to the Newtonian potential $\Phi$ via 
the lensing kernel $F_K$ through $\phi({\br})=(2/c^2)\int_0^r {\rm d}r'F_K(r,r')\Phi({\br}')$. 

\subsubsection{Intrinsic} 
The intrinsic ellipticity 3D power spectra can be written using the same formalism as the shear. For convenience we use as an example the 
linear alignment model proposed by Hirata \& Seljak (2006) where the local alignment of galaxies can be related to the primordial Newtonian 
potential
\be 
e^I(\br)=-\left(\frac{C_{\rm IA}}{2H_0^2}\right)\frac{1}{r(z)^2}\eth\eth\Phi[\br]
\ee
this is similar to the shear case, except that there is an additional amplitude $C_{\rm IA}$, 
the potential is the primordial gravitational potential, and the derivation is a comoving derivative; that we write as an  
angular part with a denominator of $r^2$, as is done in Merkel \& Schaefer (2013). 
The normalisation with $H_0^2$ ensures that the ellipticity is dimensionless. 
The potential can be linked to the density field through Poisson's equation in comoving coordinates 
$\Phi_{\ell}(k)=-4\pi G \bar\rho_{\rm M}(z) \bar D(z)^{-1} a^2 k^{-2}\delta_{\ell}(k)$, where $\bar\rho_{\rm M}(z)$ is the 
mean matter density at redshift $z$, and $\bar D(z)\propto D(z)/a$ is the normalised growth factor. This can also be written 
(see for example Hui \& Zhang, 2002) as $\Phi_{\ell}(k)=-(3H_0^2 \Omega_{\rm M}/2) \bar D(z)^{-1}a^{-1}k^{-2}\delta_{\ell}(k)$ which is the form we will use. 
Following a similar derivation to that in Kitching, Heavens, Miller (2011) the transform coefficients of the intrinsic 
ellipticity field in this model are
\ba
\label{eI}
e^I_{\ell}(k)&=& D_{I}\frac{3\Omega_{\rm M}H_0^2}{2 \pi c^2}
\int {\rm d}z_p{\rm d}z' j_{\ell}(kr[z_p])n(z_p)p(z'|z_p)\frac{1}{r[z']^2}\nn
          &&\int_0^{r[z]}d{\rm r'}\frac{I(r')\delta^D(r-r')}{a(r')}\int {\rm d}k'j_{\ell}(k' r')\frac{\delta_{\ell}(k')}{k'}, 
\ea
where we keep the pre-factor the same as equation (\ref{shear}) for comparison purposes. The kernel function is 
\be 
I[r(z)]=\left(\frac{c^2}{2 H_0^2}\right)\left(\frac{C_{\rm IA}}{\bar D(z)}\right).
\ee
This is similar to equation (\ref{shear}) except that the kernel is different, 
and only evaluated at a single comoving distance, and there is the extra $r^2$ denominator. 
These equations then need to be propagated through to the power spectra taking into account the observational aspects 
of number density and redshift distributions in a similar way to the 

We note that this expression 
has the opposite sign to the shear term, which means that the covariance between intrinsic ellipticity and shear is negative. 
Using the alternative expression for Poisson's equation leads to the same result except that the function $I$ is scaled in a different way 
giving a kernel function in equation (\ref{eI}) 
\be 
F(r[z])=\left(\frac{A_{\rm IA}}{\bar D(z)}\right)C_1\rho_{\rm crit}\Omega_{\rm M}=I(r[z])
\ee
that enters the preceding equations in a similar way to $F_K$ for the shear term. 
This is the factor of $F$ used in Heymans et al., (2013); the critical density $\rho_{\rm crit}\approx 1.4\times 10^{11} M_{\odot}$Mpc$^{-3}$, 
a normalisation $C_1\simeq 5\times 10^{-14}h^{-2}M_{\odot}^{-1}$Mpc$^3$ is required, and the parameter 
$A_{\rm IA}$ is a free parameter that is of order unity for reasonable models (see Heymans et al., 2013). 
The amplitude of the constant $C_{\rm IA}$ is expected to be very small in comparison with amplitude of the 
shear power spectrum, in order to be consistent with current observations (Joachimi et al., 2011, 2013; Mandelbaum et al., 2011). Therefore,   
and in order to link to previous studies, we adopt this scaling
\be
A_{\rm IA}=C_{\rm IA}/(C_1\rho_{\rm crit}\Omega_{\rm M})\simeq C_{\rm IA}/(2.1\times 10^{-3}),
\ee
and we will use $A_{\rm IA}$ as a free parameter in our investigations. 

We present simple changes to the $U$ and $G$ matrices that can be used to 
incorporate this into the 3D cosmic shear formalism. For the local ellipticity field we assume that only local 
potential perturbations affect the intrinsic galaxy alignment, this 
means that the kernel becomes a delta-function and the extra factor $I(r[z])$ appears
\be
\label{ui}
U^I_{\ell}(r[z],k)=\int_0^{r[z]}d{\rm r'}\frac{\delta^D(r'-r) I(r'[z])}{a(r')}j_{\ell}(kr')P^{1/2}(k; r').
\ee
This then propagates into a matrix $G$ that is similar to the shear case
\be
G^{I}_{\ell}(k_1,k')=\int {\rm d}z_p{\rm d}z' j_{\ell}(kr[z_p])n(z_p)p(z'|z_p)\frac{U^I_{\ell}(r[z'],k')}{r[z']^2},
\ee
that we will combine later with shear and CMB deflection.
\begin{figure}
 \centerline{
 \includegraphics[angle=0,width=\columnwidth]{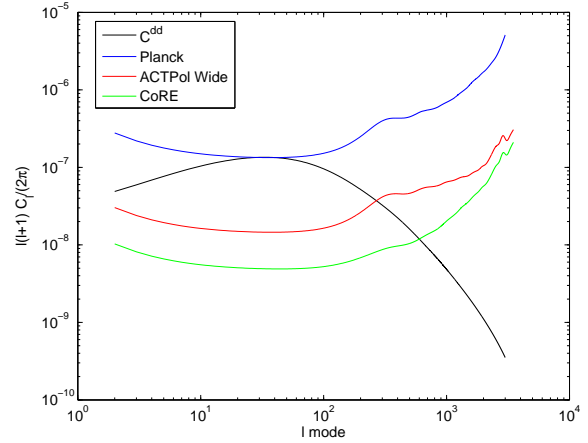}}
 \caption{The simulated reconstruction noise on the CMB deflection power spectra as a function of angular wavenumber $\ell$ for
   \emph{Planck}-like, ACTPol-like and \emph{CoRE}-like experiments, compared to the expected deflection power spectrum $C^{dd}_{\ell}$.}
 \label{cmbnoiseplot}
\end{figure}
Intrinsic alignments have been investigated within the context of 3D cosmic shear in Kitching et al., (2008) where a simple fitting formula to 
simulations was included, and in Merkel \& Schaefer (2013) who looked at II and GI effects on the 3D cosmic shear power spectrum including a quadratic 
alignment model.  

\subsubsection{CMB deflection}
To generate the terms relating the deflection field of the CMB we note that, in the continuous limit, the term 
$n(z_p)p(z'|z_p)\rightarrow \delta^D(z'-z_p)\delta^D(z'-z_{\rm CMB})$ (i.e. there is a single source plane, with negligible 
error in redshift), also that because the CMB transform is performed using a spherical harmonic transform, not a spherical-Bessel transform, 
there is no Bessel function in the associated equation for the matrix $G$ in this case. The $U$ matrix in fact remains unchanged, the lensing kernels 
functional form is the same as the shear case, except that it is only evaluated at the single redshift of the CMB. Therefore we have that 
\be 
G^{d}_{\ell}(k')=U_{\ell}(r[z_{\rm CMB}],k'). 
\ee
The $U$ matrix is unaffected in its definition, but is now an integral up to the last scattering surface only. 
The other change is that the derivative of the potential is related to the deflection field by 
\be 
d(\br)=\eth\phi(\br)
\ee
this means that the derivative terms are different for the galaxy weak lensing case, 
and we will label them here as $D_d=D_{d,1}+{\rm i}D_{d,2}=\ell_x+{\rm i}\ell_y$. 

\subsubsection{Combination}
The total shear, intrinsic and deflection power spectrum and cross-power can now be written in a compact way
\be 
\label{matall}
{\bf C}^{XY}_{\ell}={\widetilde{\bf G}^X_{\ell}}{\widetilde{\bf G}^{\dagger,Y}_{\ell}}
\ee
in the discrete case, where we show a matrix multiplication between the $G$ matrices for quantities $X$ and $Y$; $\dagger$ 
refers to a transpose and complex conjugate. The 
resulting power spectra is an $N_k\times N_k$  matrix in general. The matrix ${\widetilde{\bf G}_{\ell}}$ 
is 
\be
{\widetilde{G}^X_{\ell}(k_1,k)=D_X{\mathcal A}\frac{(\Delta k)^{1/2}}{k}G^{X}_{\ell}(k_1,k),}
\ee
where $X=\{\gamma, I, d\}$, and $\Delta k$ is the $k$-mode resolution used in the approximation of the integrals in
equation (\ref{all}) with a sum.  In the continuous $k$-mode case each $G$ matrix can be mapped to one element 
in the data vector such that a total matrix $G_{\ell}(k_1,k_2)=G^{\gamma}_{\ell}(k_1,k_2)+G^I_{\ell}(k_1,k_2)+G^d_{\ell}(k_1,k_2)$.  
In the multiplication of these within the equivalent of equation (\ref{shearpower}) the nine terms are 
\ba 
\label{all}
C^{\gamma\gamma}_{\ell}(k_1,k_2)&=&[D_{\gamma}D^*_{\gamma}]{\mathcal A}^2\int\frac{{\rm d}k'}{k'^2} G^{\gamma}_{\ell}(k_1,k')G^{\gamma}_{\ell}(k_2,k')\nn
C^{II}_{\ell}(k_1,k_2)&=&[D_{I}D^*_{I}]{\mathcal A}^2\int\frac{{\rm d}k'}{k'^2} G^I_{\ell}(k_1,k')G^I_{\ell}(k_2,k')\nn
C^{\gamma I}_{\ell}(k_1,k_2)&=&[D_{\gamma}D^*_{I}]{\mathcal A}^2\int\frac{{\rm d}k'}{k'^2} G^{\gamma}_{\ell}(k_1,k')G^I_{\ell}(k_2,k')\nn
C^{I\gamma}_{\ell}(k_1,k_2)&=&[D_{I}D^*_{\gamma}]{\mathcal A}^2\int\frac{{\rm d}k'}{k'^2} G^{I}_{\ell}(k_1,k')G^{\gamma}_{\ell}(k_2,k')\nn
C^{dd}_{\ell}&=&[D_{d}D^*_{d}]{\mathcal A}^2\int\frac{{\rm d}k'}{k'^2} G^{d}_{\ell}(k')G^{d}_{\ell}(k')\nn
C^{d\gamma}_{\ell}(k_1)&=&[D_{d}D^*_{\gamma}]{\mathcal A}^2\int\frac{{\rm d}k'}{k'^4} G^{d}_{\ell}(k')G^{\gamma}_{\ell}(k_1,k')\nn
C^{\gamma d}_{\ell}(k_1)&=&[D_{\gamma}D^*_{d}]{\mathcal A}^2\int\frac{{\rm d}k'}{k'^2} G^{\gamma}_{\ell}(k_1,k')G^{d}_{\ell}(k')\nn
C^{dI}_{\ell}(k_1)&=&[D_{d}D^*_{I}]{\mathcal A}^2\int\frac{{\rm d}k'}{k'^2} G^{d}_{\ell}(k')G^{I}_{\ell}(k_1,k')\nn
C^{Id}_{\ell}(k_1)&=&[D_{I}D^*_{d}]{\mathcal A}^2\int\frac{{\rm d}k'}{k'^2} G^{I}_{\ell}(k_1,k')G^{d}_{\ell}(k')
\ea
$D_I$ has the same $\ell$-mode dependence as $D_{\gamma}$ because both shear and the 
intrinsic ellipticity are related to 
second derivatives of potentials. Equation (\ref{all}) includes all inter-datum covariance (`cross-correlation') 
terms between the various elements in the data vector. 

The total 
ellipticity-ellipticity power spectrum, referred to in equations (\ref{pict}) and (\ref{signal}) is given by 
\ba 
\label{ee}
C^{ee}_{\ell}(k_1,k_2)&=&C^{\gamma\gamma}_{\ell}(k_1,k_2)+C^{II}_{\ell}(k_1,k_2)\nn
&+&C^{\gamma I}_{\ell}(k_1,k_2)+C^{I\gamma}_{\ell}(k_1,k_2).
\ea 
The $I\gamma$ term is expected to be zero in the absence of photometric redshift errors, because more distant intrinsic
galaxy ellipticities are not expected to be correlated with the shear from lower redshift galaxies. In the 
matrix notation presented in equation (\ref{matall}) this occurs because the $G^{\gamma}$ and $G^I$ matrices do not 
commute, the $G^{\gamma}$ matrix being approximately a Heaviside matrix in $k'$ and the $G^I$ matrix being approximately a 
delta-function matrix in $k'$ in practice. Photometric redshift errors however can cause the $I\gamma$ term to be non-zero, 
because the estimate of the source and background galaxy redshifts can be spuriously interchanged.  

Similarly the ellipticity-deflection cross-term is given by 
\be 
\label{de}
C^{de}_{\ell}(k_1)=C^{d\gamma}_{\ell}(k_1)+C^{dI}_{\ell}(k_1), 
\ee
where we include possible correlations between the intrinsic ellipticity power spectrum and the CMB deflection field. 
Each of these power spectra, through the $D$ pre-factors, have elements that can be associated with the real and imaginary parts of the shear 
field. These are combined such that the full covariance is given by the affix-covariance in equation (\ref{affix}), 
as described in Kitching et al. (2014) for both the galaxy weak lensing, CMB weak lensing and the inter-datum power spectra.

Each of the power spectra have noise terms associated with them, but the cross-power spectra do not. 
The shot-noise for the ellipticity-ellipticity power spectrum $N^{ee}_{\ell}(k_1,k_2)$ 
is given in Heavens et al., (2006), Kitching et al. (2007), and is added to 
equation (\ref{ee}). The deflection field noise term $N^{dd}_{\ell}$ is the same as that used in Das et al. (2014) that uses the 
quadratic estimator from Hu \& Okamoto (2002). When reconstructing the CMB deflection field from CMB temperature 
and/or polarization maps ($T$, $E$, $B$) one can use  
a quadratic estimator using a pair of the observables (one of $TT$, $TE$, $EB$ etc) to reconstruct the deflection field $d$. 
One then takes the power spectrum of this reconstructed $d$ field, yielding $C^{dd}_{\ell} + N^{dd}_{\ell}$ where $N^{dd}_{\ell}$ 
is the reconstruction noise. Depending on the pair XY used, one obtains the corresponding reconstruction noise, or 
one can combine the different estimators into a minimum variance one, with the noise spectrum, which is the one we use in this paper; 
we show these noise power spectra in Figure \ref{cmbnoiseplot} for the three CMB experiments described in Section \ref{Results}. 

\subsubsection{Temperature \& Polarisation Power Spectra}
For the $T$, $E$ and $B$ mode power spectra, and their covariances between each other and the deflection $d$ we use {\sc camb} to 
produce the signal. We use the noise formula provided in Taylor et al., (2006) that depends on the microwave
beam FWHM and pixel sensitivities. We refer the reader to Hu (2003) and Eisenstein et al. (1998) for a detailed explanation of these terms.  

\subsection{Galaxy Shape Measurement Systematics}
The measurement of galaxy ellipticity for weak lensing purposes 
(colloquially referred to as `shape measurement') is biased due to noise (Viola, Kitching, Joachimi, 2014); 
potential model inaccuracy, if a galaxy model-fitting approach is used (e.g. Bernstein, 2010); and algorithmic assumptions 
and errors (as quantified in the STEP and GREAT results; Heymans et al., 2006, Massey et al., 2007, Bridle et al., 2010, Kitching et al., 2012, Kitching et al., 2013). 
These biases can be parameterised by applying an additive $c$ and multiplicative $m$ bias to the inferred/observed ellipticity values such that 
$e^{\rm observed}=m e^{\rm true}+c$. We investigate the impact of multiplicative biases on the cosmological inference as a potential systematic effect.  
To include potential galaxy shape measurement systematic effects in the formalism presented in this paper 
one can simply multiply the $D_{\gamma}$ and $D_I$ factors by $m$ such 
that, for example, $D_{\gamma}\rightarrow m D_{\gamma}$ in all places that this factor appears -- note that this is in all terms in equation (\ref{ee}) 
as an $m^2$ factor, and in all terms in equation (\ref{de}) as a factor of $m$. If $m$ is redshift dependent such that $m(z)=m_0+f(z)$, where 
$f(z)$ is some function of redshift then this enters into the integral that defines the $G$ matrices. 
\ba
U_{\ell}(r[z],k)&\rightarrow&m\int_0^{r[z]}d{\rm r'}\frac{F_K(r,r')}{a(r')}j_{\ell}(k r')P^{1/2}(k; r')\nn
&+&c\left(\frac{2\pi c^2}{3\Omega_{\rm M}H_0^2}\right)\left(\frac{1}{D_{\gamma}}\right)
\ea
and similarly for the matrix $U^I_{\ell}(r[z],k)$ in equation (\ref{ui}).

In this paper we will only 
consider the redshift independent part of the multiplicative bias $m_0$ and additive bias $c_0$ for illustrative purposes.  A similar investigation was done by Das, Errard \& Spergel (2013) 
who looked at a coarsely binned 2D cosmic shear analysis, and by Vallinotto (2012, 2013). 

\section{Results} 
\label{Results}
Here we use the Fisher matrix formalism, using the covariances described in Section \ref{Methodology} 
to make predictions on the applicability of 3D cosmic shear - CMB lensing combinations to constrain 
cosmological parameters of interest. 

\subsection{Experimental Set Up}
Since we are assuming that the parameters affect
the (affix) covariance of the spherical-Bessel transform coefficients, not the mean (which is zero except for the 
effects of masks in the data), the Fisher matrix is given by
\be
\label{fish}
F_{\alpha\beta}=\frac{g}{2}\int{\rm d}\phi_{\ell}\int{\rm d}\ell\ell {\rm
  Tr}[A^{-1}_{\ell}A_{\ell},_{\alpha}A^{-1}_{\ell}A_{\ell},_{\beta}]
\ee
where we include an integral over $\ell$-space\footnote{The density of
  states accounts for correlations between modes arising from partial
  sky coverage, equivalent to the $f_{sky}$ approach of many papers.
  Note that the insensitivity to large-scale modes, which is also a
  consequence of using a patch of sky,  needs to be treated by a cut
  on $\ell$.  The Fisher matrix approach assumes the data are
  Gaussian; see  Munshi et al., (2011) for an investigation of
  non-Gaussianity in 3D cosmic shear.}
which includes a density of states in $\ell$-space, $g=A_{\rm survey}/(2\pi)^2$ where 
$A_{\rm survey}$ is the area of the survey in steradians 
(see Appendix B of Kitching et al., 2007).  A comma represents a derivative with respect to parameter
$\alpha$ or $\beta$, and the trace is over the $k$-diagonal direction in equation \ref{fish}).
\begin{figure*}
 \centerline{
 \includegraphics[angle=0,width=2.3\columnwidth]{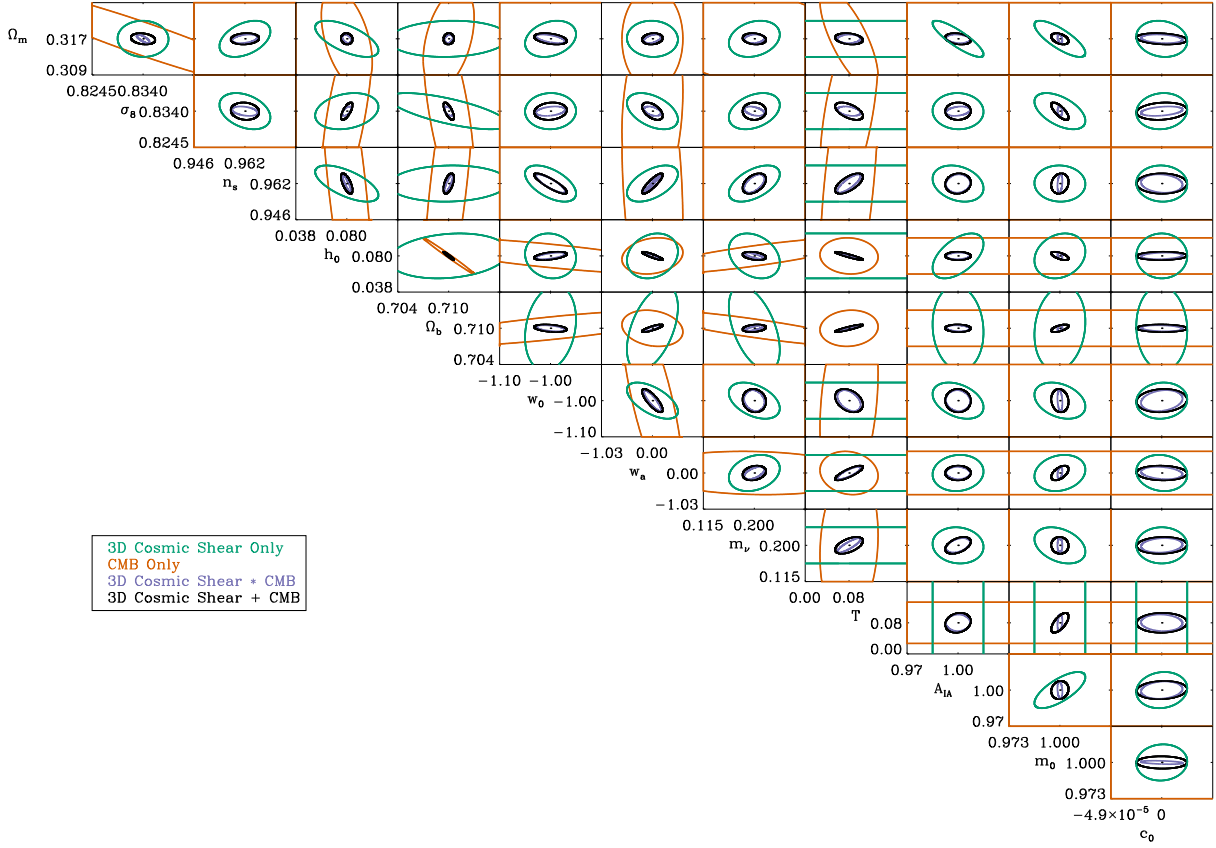}}
 \caption{The Fisher matrix $2$-parameter $1$-$\sigma$ predicted constraints for
   the case where 3D cosmic shear (for a \emph{Euclid}-like survey, green contours)
   and in combination with CMB lensing (for a \emph{Planck} survey, orange contours).
   The purple contours show the constraints including the inter-datum (``cross-correlation'')
   power spectra. The black contours show the constraints obtained by adding the 3D cosmic shear
   Fisher matrix to the CMB Fisher matrix (assuming no additional information).}
 \label{fisherplot}
\end{figure*}

Throughout we will use the parameter set (with fiducial values) :
$\Omega_m (0.3)$, $w_0(-0.95)$, $w_a(0.0)$, $h(0.71)$,
$\Omega_b(0.045)$, $\sigma_8(0.8)$, $n_s(1.0)$, $\tau(0.08)$ and for the sum of neutrino masses $m_{\nu}(0.2$eV). We use the
expansion of the dark energy equation of state as introduced in
Chevallier \& Polarski (2001), and we assume a flat geometry. We also assume that the tensor-to-scalar ratio is zero. 
For the massive neutrinos we assume a normal hierarchy (see Jimenez et al., 2010 for a discussion of 
how this assumption affects expected error and evidence predictions). Additional parameters $A_{\rm IA}(1.0)$, $m_0(1.0)$ 
and $c_0(0.0)$, parameterise galaxy weak 
lensing systematic effects for intrinsic alignments and galaxy shape measurement respectively as described in Section 
\ref{Methodology}.

We investigate near-term and longer-term 3D cosmic shear survey configurations of $1500$ and 
$15$,$000$ square degrees respectively, with a surface number densities of $15$ and $30$ galaxies per square
arcminutes, and an intrinsic ellipticity dispersion of $0.3$. Where we
do not use direct photometric redshift probabilities we will use a
redshift distribution of galaxies $n(z)\propto z^2{\rm exp}[-(1.4z/z_m)^{1.5}]$ 
with a median redshift of $z_m=1$ and a Gaussian redshift dispersion
with a redshift error $\sigma_z(z)=0.03(1+z)$. These survey configurations
are similar to the ESO KiDS survey (de Jong et al., 2013) 
and the ESA \emph{Euclid}\footnote{{\tt http://euclid-ec.org}} wide survey (Laureijs et al., 2011). 
Throughout we use a maximum radial wavenumber of either $k_{\rm max}=1.5h$Mpc$^{-1}$ or $k_{\rm max}=5.0h$Mpc$^{-1}$ 
to investigate the scale-dependence of the results and to avoid the highly non-linear regime $k_{\rm max}>5.0h$Mpc$^{-1}$ 
where theoretical predictions for the power spectrum may be unsound, however baryonic effects persist to lower $k$ (e.g.
White, 2004; Zhan \& Knox, 2004; Jing et al., 2006; Zenter et al., 2008; 
Kitching \& Taylor, 2011; van Daalen et al., 2011; Semboloni et al. 2011, 2013). These maximum $k$-mode values are conservative with 
respect to those used in correlation function analyses (e.g. Heymans et al., 2013), however MacCrann et al. (2014) claim such 
analyses are not sensitive to small-scale baryonic effects. In particular, the intrinsic alignment model is more uncertain at 
small scales, we refer the reader to Troxel \& Ishak (2014) and references therein for a discussion 
of this point and the possibility of improvement using halo modelling. The $k$-mode cuts we use imply an effective azimuthal $\ell$-mode cut of approximately 
$\ell_{\rm max}\simeq 5000$ through the Bessel function behaviour $j_{\ell}(kr)\simeq 0$ for $\ell\gs kr$ where $r$ is a comoving distance.    
\begin{table*}
\begin{tabular}{|c|l|l|l|l|}
Parameter& 3D Cosmic Shear Only&CMB Only&3D Cosmic Shear+CMB&3D Cosmic Shear$*$CMB\\
\hline
\hline
Comological Paramaters\\
$\Omega_{\rm M}$&          $0.0041$ &  $0.0112$ & $0.0012$ & $0.0009$\\
$\sigma_8$      & $0.0047$   &$ 0.0258$ & $0.0022$ & $ 0.0012$\\
$n_s$     &$0.0081$  &$0.0718$ &$0.0045$ &$0.0041$\\
$h$       &$0.0262$ &$0.0211$ &$0.0047$ &$0.0038$\\
$\Omega_{\rm B}$  &$0.0068$ &$0.0029$ &$0.0006$ &$0.0005$\\
$w_0$     &$0.0490$   &$0.2665$ &$0.0311$  &$0.0282$\\
$w_a$    &$0.5167$  &$0.6179$  &$0.2078$ &$0.1758$\\
$m_{\nu}$ &$0.0425$ &$0.1824$ &$0.0195$ & $0.0170$\\
$\tau$ &     &    $0.0530$   &$0.0253$ & $0.0213$\\
\hline
Galaxy Systematic Effect Paramaters\\
$A_{\rm IA}$   &$0.0157$  & &    $0.0078$  &$0.0075$\\
$m_0$        &$0.0135$ & &  $0.0046$ &$0.0013$\\
$c_0$       &$0.00002$ & &  $0.00002$  &$0.00002$\\
\hline
FoM &      $49$     & $9$ &     $254$   &   $409$\\
\hline
FoM with no systematics effects & $55$ & $9$ &$376$&$448$\\
\hline
\end{tabular}
\caption{The predicted $1$-sigma 
marginalised errors on the cosmological and systematic parameters for the 3D cosmic shear only case (for a \emph{Euclid}-like survey) 
and for a CMB only case (\emph{Planck}), and in combination by assuming no inter-datum information 
(denoted by $+$) and with the inter-datum information
included (denoted by $*$). Also shown is the dark energy Figure of Merit (FoM), and also the FoM where we do not marginalise over any of the 
galaxy lensing systematic effects.}
\label{table}
\end{table*}

We investigate three CMB experiments \emph{Planck} (Planck Collaboration, 2006), ACTPoL (Niemack et al., 2010) and for a possible large angular-scale 
polarisation satellite mission we use the \emph{COrE} (COrE Collaboration et al., 2011) specifications. 
For \emph{Planck} we use the temperature and polarisation 
sensitivities given by the Planck Collaboration (2006). For ACTPoL we use the temperature and polarisation sensitivities given by Niemack et al., (2010). 
For all CMB surveys we assume complete overlapping sky coverage with both of the imaging surveys considered. 
For the CMB experiments we use a maximum azimuthal wavenumber of $\ell_{\rm max}=3000$. 
In the case of ACTPoL we also assume that \emph{Planck} data is available, and so supplement the ACTPoL bands with the \emph{Planck} bands. 
We present Fisher matrix results for all experiments\footnote{We use a two-step derivative in the Fisher matrix calculation with a step size of $10\%$ of the 
fiducial parameter value (or $0.1$ if that value is zero). We tested numerical 
stability by using multiple step sizes, for both one and two-step derivatives. These tests show that expected errors are accurate to better than $2\%$ 
of their quoted values.} (despite the fact that \emph{Planck} already has temperature data published) such that 
a fair comparison can be made, and also so that we can include expected \emph{Planck} polarisation measurements. 

\subsection{Parameter Results}
In Table \ref{table} and Figure \ref{fisherplot} we show the predicted cosmological parameter constraints for a \emph{Euclid}-like galaxy weak lensing 
survey combined with a \emph{Planck} CMB survey. We show results taking into account the full covariance between the experiments, and also results 
assuming that such inter-datum covariance is zero (a simple addition of the individual Fisher matrices). We find that for 
the additional inter-datum covariance does not add significantly new information but decreases all error bars by a small amount, 
however for three exceptions, $w_0$, $w_a$ and $m_{\nu}$, there is a notable 
reduction in the predicted parameter error. 
Even in the case that the extra inter-datum covariance does not improve the predicted constraints, the addition
of the CMB information still improves the predicted parameter constraints through the intersection of the parameter confidence ellipsoids.

For the dark energy parameters, $w_0$ and $w_a$, we find that the extra information can reduce marginalised error bars by $15\%$, and 
the inverse of the area of the projected $(w_0,w_a)$ confidence ellipse, parameterised by the dark energy 
`Figure of Merit' (Albrecht et al., 2006) increases by similar factors. This is because the inter-datum covariance improves measurements 
of the expansion history and integrated growth of structure from the CMB over the redshift 
range of the galaxy weak lensing survey. A similar improvement was found in Vallinotto (2012, 2013).
In general it should be expected that any parameter that strongly affects 
amplitude changes in the the redshift evolution of the power spectra will be measured more accurately by including the inter-datum covariance. 
However, the methodology used for intrinsic alignment modelling is currently uncertain (see Troxel \& Ishak, 2014 for a review), 
and the detailed numerical results here will depend on the model used.

\begin{figure*}
 {\bf Euclid-like \& Planck}\\
 \includegraphics[angle=0,width=1.5\columnwidth,clip=]{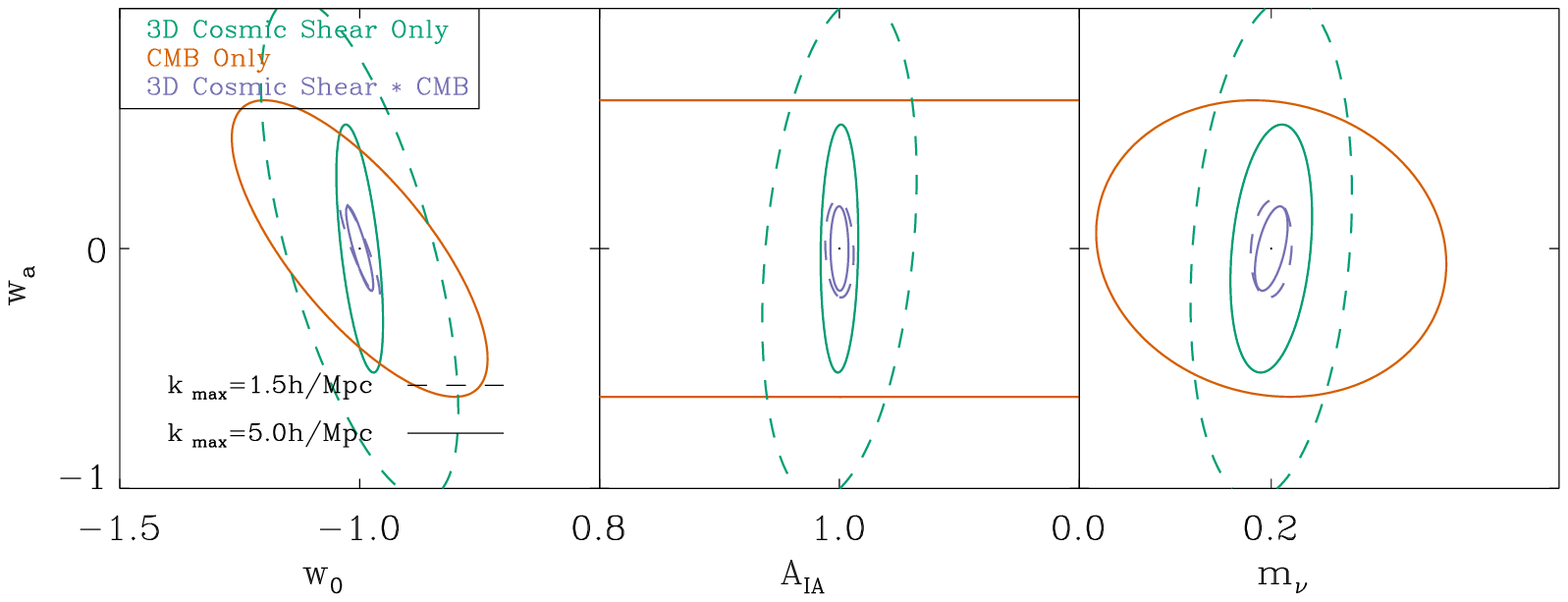}\\
 {\bf Euclid-like \& Planck+ACTPoL}\\
 \includegraphics[angle=0,width=1.5\columnwidth,clip=]{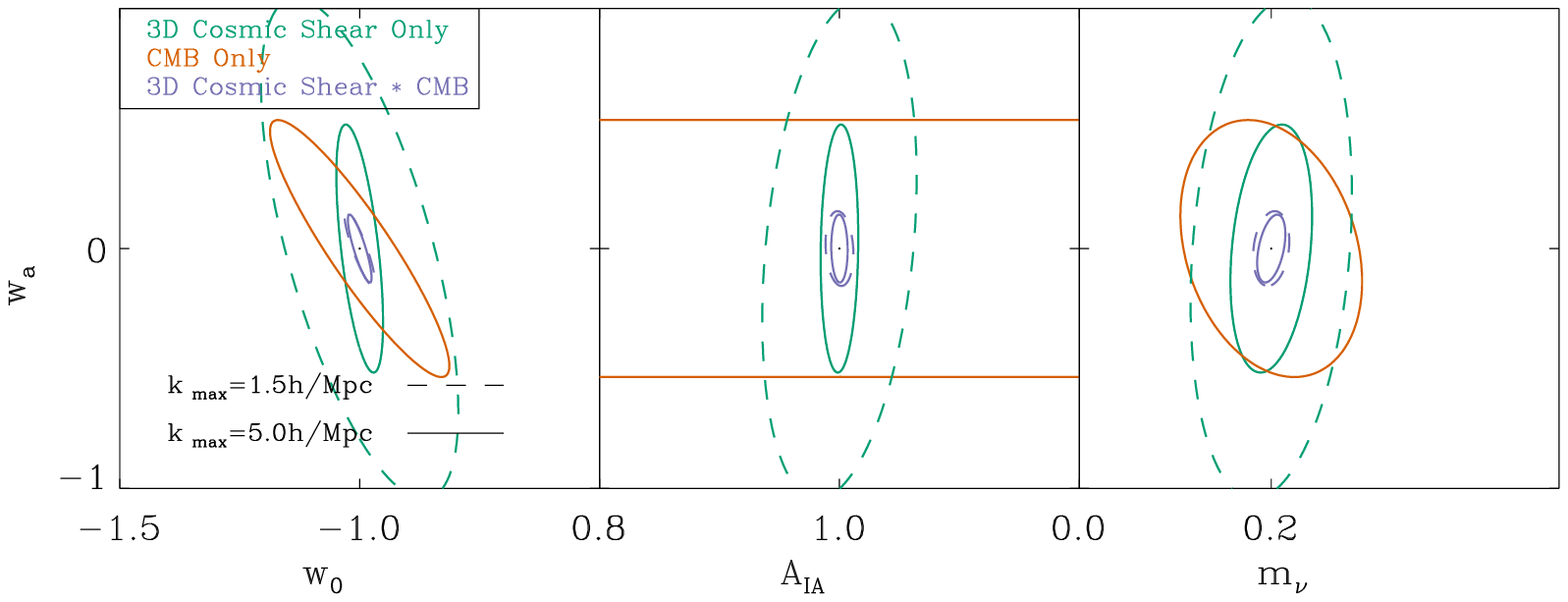}\\
 {\bf Euclid-like \& COrE-like}\\
 \includegraphics[angle=0,width=1.5\columnwidth,clip=]{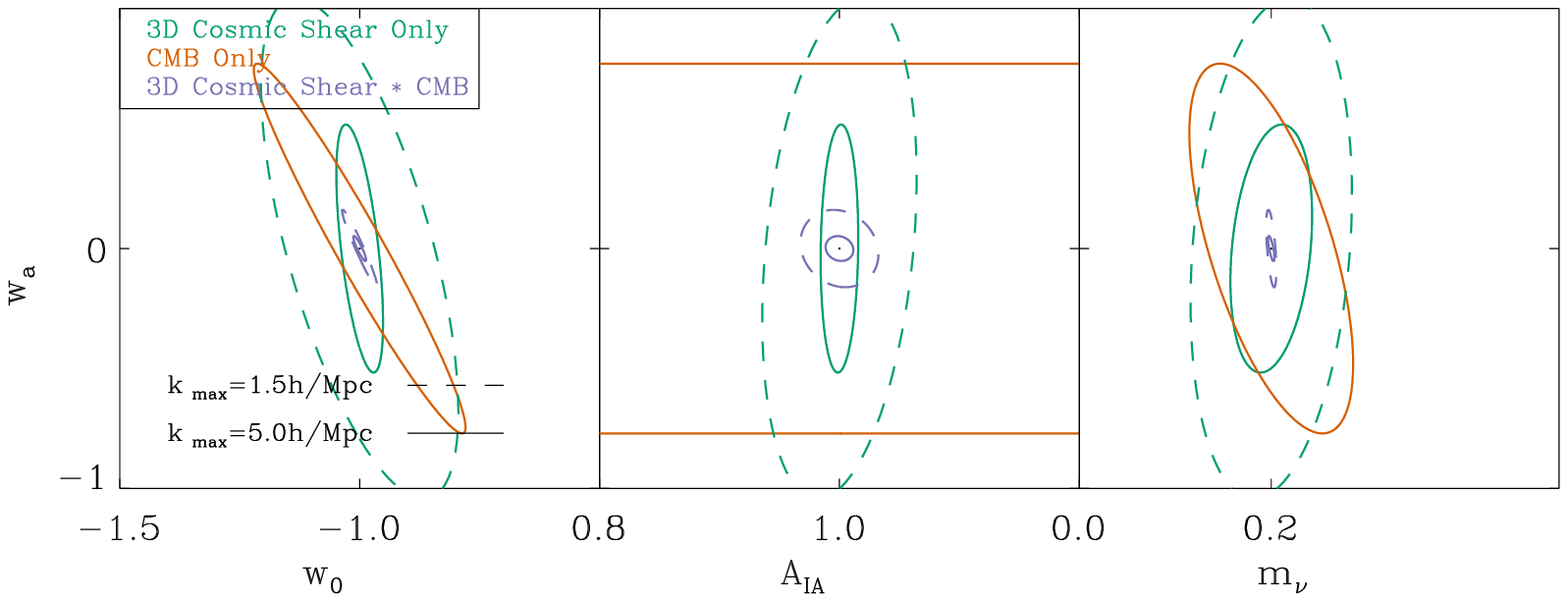}\\
 {\bf KiDS \& Planck+ACTPoL}\\
 \includegraphics[angle=0,width=1.5\columnwidth,clip=]{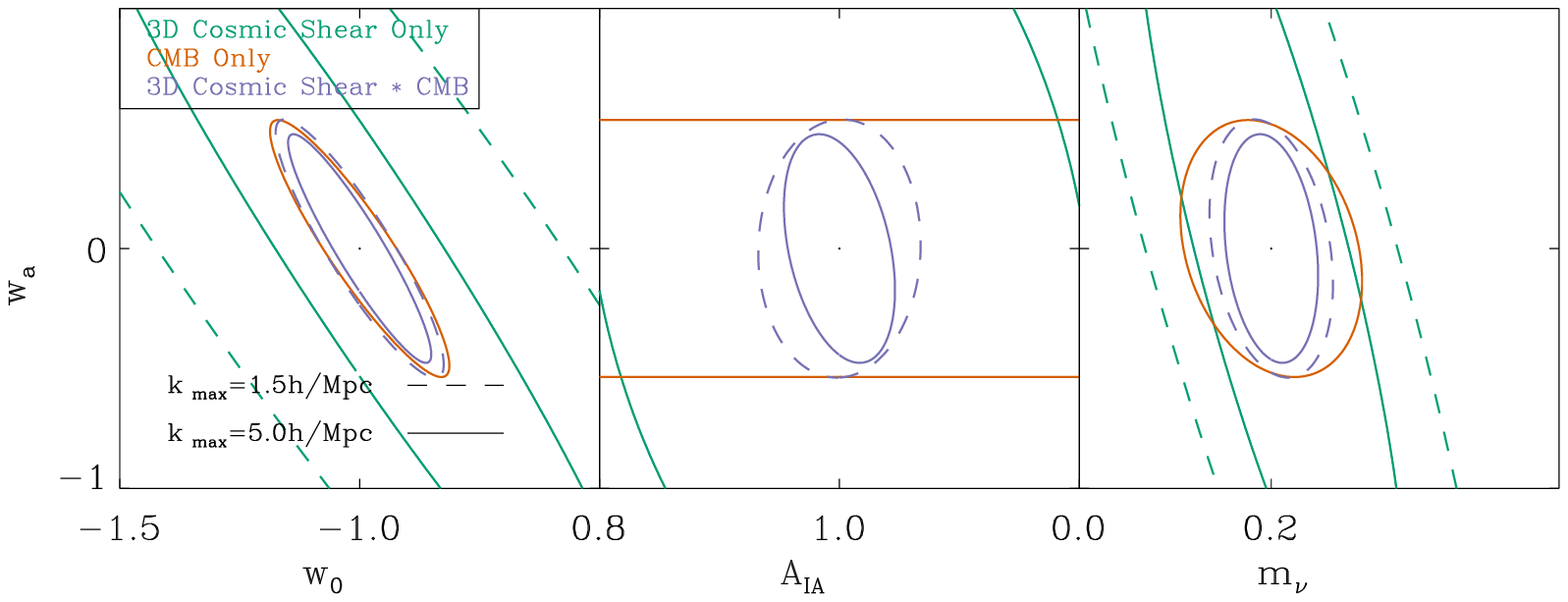}
 \caption{The Fisher matrix $2$-parameter $1$-$\sigma$ predicted constraints for
   the projected $(w_a,w_0)$, $(w_a, A_{\rm IA})$ and $(w_a, m_{\nu})$ parameter spaces. 
   The purple contours show the contraints including the inter-datum (``cross-correlation'')
   power spectra. The green and orange contours show the 3D cosmic shear and CMB constraints respectively. The dashed and solid 
   contours use a maximum radial in the 3D cosmic shear calculation of $k_{\rm max}=1.5h$Mpc$^{-1}$ 
   and $k_{\rm max}=5.0h$Mpc$^{-1}$ respectively.}
 \label{fisherplot2}
\end{figure*}

In Figure \ref{fisherplot2} we show the predicted marginalised constraints for the parameters for which we find an improvement when 
including the inter-datum covariance between 3D cosmic shear and the CMB, for a maximum radial wavenumber included in the 3D cosmic shear analysis of 
$k_{\rm max}=1.5h$Mpc$^{-1}$ and $k_{\rm max}=5.0h$Mpc$^{-1}$. We show the combination of a \emph{Euclid}-like survey with the three CMB experiments considered, and also the combination 
of KiDS with ACTPoL. It is clear from this Figure that the removal of scales between $1.5< k < 5.0h$Mpc$^{-1}$ significantly degrades the dark energy Figure of Merit for 
3D cosmic shear - a change for a \emph{Euclid}-like survey of over a factor of ten from $\sim 50$ to $5$ -- however that the combination of CMB allows for a recovery of the information, reaching values of approximately $\sim 300$ including \emph{Planck} using only $k_{\rm max}\leq 1.5h$Mpc$^{-1}$. 
This means in principle 
that poorly understood non-linear scales in 3D cosmic shear could be removed and that the overall dark energy science, 
in combination with CMB information, could be recovered: a `clean' cosmological probe. We also show predictions for the KiDS survey and show that intrinsic alignments can be calibrated in such a survey using CMB information from ACTPoL, even in the case that only linear scales are used. When non-linear scales are included KIDS is expected to 
improve dark energy measurements from the CMB alone by a factor of two -- a change in dark energy Figure of Merit from $\sim 25$ for ACTPoL alone 
to $\sim 40$ with KiDS included. 

A \emph{Euclid}-like 3D cosmic shear experiment in combination with the expected performance from \emph{COrE} results in a significant improvement in the dark energy Figure of Merit from $\sim 400$ for \emph{Planck} to $\sim 3000$ with \emph{COrE}. 
We see a similar improvement for the sum of neutrino mass constraints, when
combining a \emph{Euclid}-like 3D cosmic shear experiment with either 
\emph{Planck} or ACTPoL results in errors of $\sim 0.015$ eV, whereas in combination with \emph{COrE} this is a factor of $5$ times 
smaller with an expected error of 
$\ls 0.003$ eV. Hall \& Challinor (2012) found similar expected errors for a \emph{CoRE}-like experiment 
using MCMC methods, and we also find similar expected errors to Kitching et al. (2007) who considered 
slightly different 3D cosmic shear survey characteristics and did not use the advances described in Kitching et al. (2014). 
This small expected error on the sum of the neutrino masses raises the possibility that errors on the masses of the individual neutrino species may be small enough 
to determine the neutrino hierarchy as discussed in Jimenez et al. (2010), although this is likely 
only if there is a normal hierarchy, with close-to-minimal mass, 
as shown in Hamann et al. (2012).

\section{Conclusion}
\label{Conclusion}
The current best probe of cosmology is the CMB, observations of which have helped to define the current cosmological model. 
However to determine 
the nature of the dominant components of that model, dark energy and dark matter, requires new cosmological probes and galaxy weak lensing 
combined with galaxy redshift information - 3D cosmic shear - is one such probe. The CMB is weakly 
gravitationally lensed by large-scale structure along the line of sight, and galaxy images are also weakly lensed by large-scale structure. Therefore
in order to correctly combine these two data sets requires the calculation of the covariance between them.

In this paper we have shown how CMB and 3D cosmic shear data can be combined in a self-consistent spherical-Bessel power spectrum 
statistic. We include the inter-datum covariance (`cross-correlation') 
between the CMB and 3D cosmic shear in this formalism. 
We also include galaxy intrinsic alignments and galaxy shape measurement errors, including the full covariance between galaxy ellipticity and CMB weak lensing deflection.  

We find that the inclusion of the inter-datum covariance improves parameter constraints in particular on the dark energy equation of state evolution, and 
on the amplitude of galaxy intrinsic alignments. We find that the expected error on the linear-alignment amplitude in galaxy weak lensing can be 
improved by a factor of two by correctly including CMB information. 

By including CMB information as a baseline cosmological probe 3D cosmic shear surveys are likely 
to be able to calibrate simple intrinsic alignment models and shape measurement systematics, 
by including a small number of nuisance parameters, and still achieve their dark energy science objectives. 

\section*{Acknowledgments}
TDK is supported by a Royal Society University Research Fellowship. We thank Anthony Challinor, Jan Hamann, Alberto Vallinotto for useful comments. 
We especially thank the referee, Alex Hall, for a very careful and detailed referee report that has improved the manuscript significantly.


\begin{thebibliography}{99}

\bibitem[Albrecht et al.(2006)]{2006astro.ph..9591A} Albrecht, A., Bernstein, G., Cahn, R., et al.\ 2006, arXiv:astro-ph/0609591

\bibitem[Bridle et al.(2010)]{2010MNRAS.405.2044B} Bridle, S., Balan, S.~T., Bethge, M., et al.\ 2010, \mnras, 405, 2044 

\bibitem[Castro et al.(2005)]{2005PhRvD..72b3516C} Castro, P.~G., Heavens, 
A.~F., \& Kitching, T.~D.\ 2005, PRD, 72, 023516 

\bibitem{Chevallier} Chevallier, Polarski D., 2001, JMPD, 10, 213

\bibitem[The COrE Collaboration et al.(2011)]{2011arXiv1102.2181T} The COrE Collaboration, Armitage-Caplan, C., Avillez, M., et al.\ 2011, arXiv:1102.2181 

\bibitem[Cropper et al.(2013)]{2013MNRAS.431.3103C} Cropper, M., Hoekstra, H., Kitching, T., et al.\ 2013, \mnras, 431, 3103 

\bibitem[Das et al.(2013)]{2013arXiv1311.2338D} Das, S., Errard, J., \& Spergel, D.\ 2013, arXiv:1311.2338

\bibitem[Das et al.(2014)]{2014JCAP...04..014D} Das, S., Louis, T., Nolta, M.~R., et al.\ 2014, jcap, 4, 14 

\bibitem[Eisenstein et al.(1999)]{1999ApJ...518....2E} Eisenstein, D.~J., Hu, W., \& Tegmark, M.\ 1999, \apj, 518, 2 

\bibitem[Hall \& Challinor(2014)]{2014arXiv1407.5135H} Hall, A., \& Challinor, A.\ 2014, arXiv:1407.5135

\bibitem[Hall \& Taylor(2014)]{2014arXiv1401.6018H} Hall, A., \& Taylor, A.\ 2014, arXiv:1401.6018

\bibitem[Hamann et al.(2012)]{2012JCAP...11..052H} Hamann, J., Hannestad, S., \& Wong, Y.~Y.~Y.\ 2012, jcap, 11, 52 

\bibitem[Hand et al.(2013)]{2013arXiv1311.6200H} Hand, N., Leauthaud, A., Das, S., et al.\ 2013, arXiv:1311.6200 

\bibitem{Heavens03} Heavens A.F., 2003, MNRAS, 343, 1327

\bibitem[Heavens et al.(2013)]{2013MNRAS.433L...6H} Heavens, A., Alsing, J., \& Jaffe, A.~H.\ 2013, \mnras, 433, L6 

\bibitem{Heavens06} Heavens A.F., Kitching T.D., Taylor A.N., 2006, MNRAS, 373, 105

\bibitem[Heavens et al.(2007)]{2007MNRAS.380.1029H} Heavens, A.~F., Kitching, T.~D., \& Verde, L.\ 2007, \mnras, 380, 1029 

\bibitem[Heymans et al.(2006)]{2006MNRAS.368.1323H} Heymans, C., Van Waerbeke, L., Bacon, D., et al.\ 2006, \mnras, 368, 1323 

\bibitem[Heymans et al.(2013)]{2013MNRAS.432.2433H} Heymans, C., Grocutt, E., Heavens, A., et al.\ 2013, \mnras, 432, 2433 

\bibitem[Hirata \& Seljak(2004)]{2004PhRvD..70f3526H} Hirata, C.~M., \& Seljak, U.\ 2004, \prd, 70, 063526 

\bibitem[Hu(2003)]{2003PhRvD..67h1304H} Hu, W.\ 2003, \prd, 67, 081304 

\bibitem[Hu \& Okamoto(2002)]{2002ApJ...574..566H} Hu, W., \& Okamoto, T.\ 2002, \apj, 574, 566 

\bibitem[Hui \& Zhang(2002)]{2002astro.ph..5512H} Hui, L., \& Zhang, J.\ 2002, arXiv:astro-ph/0205512 

\bibitem[Jimenez et al.(2010)]{2010JCAP...05..035J} Jimenez, R., Kitching, T., Pe{\~n}a-Garay, C., \& Verde, L.\ 2010, jcap, 5, 35 

\bibitem[Jing et al.(2006)]{2006ApJ...640L.119J} Jing, Y.~P., Zhang, P., Lin, W.~P., Gao, L., \& Springel, V.\ 2006, \apjl, 640, L119

\bibitem[Joachimi et al.(2011)]{2011A&A...527A..26J} Joachimi, B., Mandelbaum, R., Abdalla, F.~B., \& Bridle, S.~L.\ 2011, AAP, 527, A26

\bibitem[Joachimi et al.(2013)]{2013MNRAS.431..477J} Joachimi, B., Semboloni, E., Bett, P.~E., et al.\ 2013, \mnras, 431, 477

\bibitem[de Jong et al.(2013)]{2013ExA....35...25D} de Jong, J.~T.~A., Verdoes Kleijn, G.~A., Kuijken, K.~H., \& Valentijn, E.~A.\ 2013, Experimental Astronomy, 35, 25 

\bibitem[Kitching (2007)]{} Kitching, T.~D; PhD Thesis, University of Edinburgh \ 2007

\bibitem[Kitching et al.(2012)]{2012MNRAS.423.3163K} Kitching, T.~D., Balan, S.~T., Bridle, S., et al.\ 2012, \mnras, 423, 3163

\bibitem[Kitching et al.(2014)]{2014MNRAS.442.1326K} Kitching, T.~D., Heavens, A.~F., Alsing, J., et al.\ 2014, \mnras, 442, 1326 

\bibitem[Kitching et al.(2011)]{2011MNRAS.413.2923K} Kitching, T.~D., Heavens, A.~F., \& Miller, L.\ 2011, \mnras, 413, 2923 

\bibitem[Kitching et al.(2007)]{2007MNRAS.376..771K} Kitching, T.~D., Heavens, A.~F., Taylor, A.~N., et al.\ 2007, \mnras, 376, 771 

\bibitem[Kitching et al.(2013)]{2013ApJS..205...12K} Kitching, T.~D., Rowe,  B., Gill, M., et al.\ 2013, \apjs, 205, 12

\bibitem[Kitching \& Taylor(2011)]{2011MNRAS.416.1717K} Kitching, T.~D., \& Taylor, A.~N.\ 2011, \mnras, 416, 1717 

\bibitem[Laureijs et al.(2011)]{2011arXiv1110.3193L} Laureijs, R., Amiaux, J., Arduini, S., et al.\ 2011, arXiv:1110.3193

\bibitem[Lewis \& Challinor(2006)]{2006PhR...429....1L} Lewis, A., \& Challinor, A.\ 2006, \physrep, 429, 1 

\bibitem[Niemack et al.(2010)]{2010SPIE.7741E..1SN} Niemack, M.~D., Ade, P.~A.~R., Aguirre, J., et al.\ 2010, \procspie, 7741,

\bibitem[MacCrann et al.(2014)]{2014arXiv1408.4742M} MacCrann, N., Zuntz, J., Bridle, S., Jain, B., \& Becker, M.~R.\ 2014, arXiv:1408.4742 

\bibitem[Mandelbaum et al.(2011)]{2011MNRAS.410..844M} Mandelbaum, R., Blake, C., Bridle, S., et al.\ 2011, \mnras, 410, 844

\bibitem[Massey et al.(2007)]{2007MNRAS.376...13M} Massey, R., Heymans, C., Berg{\'e}, J., et al.\ 2007, \mnras, 376, 13 

\bibitem[Merkel \& Sch{\"a}fer(2013)]{2013MNRAS.434.1808M} Merkel, P.~M., \& Sch{\"a}fer, B.~M.\ 2013, \mnras, 434, 1808 

\bibitem[Munshi et al.(2011)]{2011MNRAS.416.1629M} Munshi, D., Kitching, T., Heavens, A., \& Coles, P.\ 2011, \mnras, 416, 1629 

\bibitem[The Planck Collaboration(2006)]{2006astro.ph..4069T} Planck Collaboration 2006, arXiv:astro-ph/0604069 

\bibitem[Planck Collaboration et al.(2013)]{2013arXiv1303.5076P} Planck Collaboration 2013a, arXiv:1303.5076 

\bibitem[Planck Collaboration et al.(2013)]{2013arXiv1303.5077P} Planck Collaboration, 2013b, arXiv:1303.5077 

\bibitem[Semboloni et al.(2013)]{2013MNRAS.434..148S} Semboloni, E., Hoekstra, H., \& Schaye, J.\ 2013, \mnras, 434, 148

\bibitem[Semboloni et al.(2011)]{2011MNRAS.417.2020S} Semboloni, E., Hoekstra, H., Schaye, J., van Daalen, M.~P., \& McCarthy, I.~G.\ 2011, \mnras, 417, 2020

\bibitem[Taylor et al.(2007)]{2007MNRAS.374.1377T} Taylor, A.~N., Kitching, T.~D., Bacon, D.~J., \& Heavens, A.~F.\ 2007, \mnras, 374, 1377

\bibitem[Troxel \& Ishak(2014)]{2014arXiv1407.6990T} Troxel, M.~A., \& Ishak, M.\ 2014a, arXiv:1407.6990 

\bibitem[Troxel \& Ishak(2014)]{2014PhRvD..89f3528T} Troxel, M.~A., \& Ishak, M.\ 2014b, \prd, 89, 063528 

\bibitem[van Daalen et al.(2011)]{2011MNRAS.415.3649V} van Daalen, M.~P., Schaye, J., Booth, C.~M., \& Dalla Vecchia, C.\ 2011, \mnras, 415, 3649

\bibitem[van Engelen et al.(2014)]{2014ApJ...786...13V} van Engelen, A., Bhattacharya, S., Sehgal, N., et al.\ 2014, \apj, 786, 13 

\bibitem[Vallinotto(2012)]{2012ApJ...759...32V} Vallinotto, A.\ 2012, ApJ, 759, 32

\bibitem[Vallinotto(2013)]{2013ApJ...778..108V} Vallinotto, A.\ 2013, ApJ, 778, 108 

\bibitem[Viola et al.(2014)]{2014MNRAS.439.1909V} Viola, M., Kitching, T.~D., \& Joachimi, B.\ 2014, \mnras, 439, 1909 

\bibitem[White(2004)]{2004APh....22..211W} White, M.\ 2004, Astroparticle Physics, 22, 211

\bibitem[Zaldarriaga \& Seljak(1997)]{1997PhRvD..55.1830Z} Zaldarriaga, M., \& Seljak, U.\ 1997, \prd, 55, 1830 

\bibitem[Zentner et al.(2008)]{2008PhRvD..77d3507Z} Zentner, A.~R., Rudd, D.~H., \& Hu, W.\ 2008, \prd, 77, 043507

\bibitem[Zhan \& Knox(2004)]{2004ApJ...616L..75Z} Zhan, H., \& Knox, L.\ 2004, \apjl, 616, L75


\end{thebibliography}
\end{document}